\documentstyle[12pt]{article}

\catcode `\@=11
\if@twoside
\else
\fi
\def\section{\@startsection {section}{1}{\z@}{3.5ex plus 1ex minus
 .2ex}{2.3ex plus .2ex}{\normalsize\bf}}
\def\subsection{\@startsection{subsection}{2}{\z@}{3.25ex plus 1ex %
minus
 .2ex}{1.5ex plus .2ex}{\normalsize\sl}}
\def\subsubsection{\@startsection{subsubsection}{3}{\z@}{3.25ex plus
 1ex minus .2ex}{1.5ex plus .2ex}{\normalsize\it}}

\newif\ifrunhead

\def\title{
 \ifrunhead \vspace{0.5cm}\vspace{1ex} \fi
 \vspace{4ex}
 \bgroup
 \obeylines
 \large\boldmath \bf\begin{center}
}
\def\endtitle{\end{center}\vskip1sp\egroup}
\def\author#1{\bgroup\center #1 \endcenter\egroup}
\def\moreauthors#1{\hbox to\textwidth{\hss\vrule height.8cm %
width0pt\relax
#1\hss}}
\def\instit{\bgroup\small\it\obeylines\center}
\def\endinstit{\endcenter\vskip1sp\egroup}
\def\receipt#1{\bgroup\small\center
(Received #1)
\endcenter\egroup}
\def\abstract{\bgroup\vskip1sp \quotation}
\def\endabstract{\vskip1sp\endquotation\egroup}

\newcounter{figcaption}
\def\thefigcaption{\arabic{figcaption}}
\def\fnum@figcaption{{\bf Fig. \thefigcaption :}}
\def\figcaption{
 \par\pagebreak {\parindent 0pt \bf Figure Captions} \par %
\vskip 10pt
 \list{\fnum@figcaption}
 {\leftmargin 5em \labelwidth\leftmargin\advance\labelwidth-\labelsep
 \def\makelabel##1{##1\hfil} \usecounter{figcaption}}
}

\def\references#1{\section*{References\@mkboth
 {REFERENCES}{REFERENCES}}\list
 { \arabic{enumi})\ }{\settowidth\labelwidth{#1)\ %
}\leftmargin\labelwidth
 \advance\leftmargin\labelsep \usecounter{enumi}}
 \def\newblock{\hskip .11em plus .33em minus -.07em}
 \sloppy \sfcode`\.=1000\relax}

\def\cite{\@ifnextchar [{\@tempswatrue\@citex}{%
\@tempswafalse\@citex[]}}
\def\@citex[#1]#2{\if@filesw\immediate\write\@auxout{%
\string\citation{#2}}\fi
 \def\@citea{}\@cite{\@for\@citeb:=#2\do
 {\if-\@citeb \mbox{-}\def\@citea{}
 \else
 \@citea\def\@citea{,\penalty\@m}\@ifundefined
 {b@\@citeb}{{\bf ?}\@warning
 {Citation `\@citeb' on page \thepage \space undefined}}
 \hbox{\csname b@\@citeb\endcsname}
 \fi}}{#1}}
\newfont{\scrptrm}{cmr8}
\def\@cite#1#2{$[{#1\if@tempswa , #2\fi}]$}
\def\refcite{\@ifnextchar [{\@tempswatrue\@refcitex}
 {\@tempswafalse\@refcitex[]}}
\def\@refcitex[#1]#2{
 \if@filesw\immediate\write\@auxout{\string\citation{#2}}\fi
 \def\@citea{}\@refcite{\@for\@citeb:=#2\do
 {\if-\@citeb -\def\@citea{}
 \else
 \@citea\def\@citea{,\penalty\@m}\@ifundefined
 {b@\@citeb}{{\bf ?}\@warning
 {Citation `\@citeb' on page \thepage \space undefined}}
 \hbox{\csname b@\@citeb\endcsname}
 \fi}}{#1}}
\def\@refcite#1#2{{#1\if@tempswa , #2\fi}}

\catcode `\@=12

\newcommand{\beq}{\begin{equation}}
\newcommand{\eeq}{\end{equation}}
\newcommand{\beqa}{\begin{eqnarray}}
\newcommand{\eeqa}{\end{eqnarray}}

\begin{document}

\begin{center}

\title
  Diffusion of electrons in random magnetic fields
\endtitle
\author{Tohru Kawarabayashi and Tomi Ohtsuki$^*$}
\instit
Institute for Solid State Physics, University of Tokyo,
Roppongi, Minato-ku, Tokyo 106, Japan  \\
${}^*$Department of Physics, %
Faculty of Science, Toho University, Miyama 2-2-1,
Funabashi 274, Japan
\endinstit

\vskip 0.4in

Abstract
\end{center}

\vspace*{1cm}

Diffusion of electrons in a two-dimensional system
in static random magnetic fields
is studied by solving the time-dependent %
Schr\"{o}dinger equation
numerically. The
asymptotic behaviors of the second moment of %
the wave packets and
the temporal
auto-correlation function in such systems
are investigated. It is shown that, in the region away from
the band edge,
the growth of the second moment of the wave packets turns out to be
diffusive, whereas
the exponents for the power-law decay of the %
temporal auto-correlation
function
suggest a kind of fractal structure %
in the energy spectrum and in the wave
functions.
The present results are consistent with %
the interpretation that the
states away from the band edge region are critical.

\vspace*{1cm}


\newpage
%









%
\section{Introduction}

Since the development of the scaling theory of localization
\cite{AALR,Kawabata}, much
work has been done on the transport
properties of disordered electron
systems \cite{LR,edKS}.
To understand the transport properties of a random system, the
concept of quantum interference plays an important role. In fact,
in the last decade, many interesting and universal
phenomena observed in
mesoscopic systems have been understood in terms of
the quantum interference \cite{edKramer,KramerMacKinnon}.
In most cases, however, only a coherent scattering
due to the random potentials has been discussed. Recently
there has been a growing interest
on the coherent scattering by the randomly
varying
magnetic fields\cite{AI,-,Ohtsukietal}.
In particular, considerable attention
has been  paid to transport properties in two-dimensional
systems in randomly distributed magnetic fluxes. For instance,
several authors \cite{PZ,-,LXSZ} have
discussed  whether or not
the Anderson transition occurs in a two-dimensional
system in static random magnetic fields.
A possibility of the Anderson transition in such a system
has been proposed in
references \cite{AHK}, \cite{KWAZ} and \cite{LXSZ}
based on numerical calculations of conductance and
the scaling analysis for the exponential-decay length of the
Green function in quasi-one-dimensional systems.
According to the references
\cite{AHK} and \cite{KWAZ}, the mobility
edge $E_c$ has been estimated
as $E_c = 2.6 \sim 3.0$. On the other hand,
there exist numerical data which  suggest
the absence of such a transition \cite{SN}.
It has been also argued quite recently that
the model in question
can be mapped to the nonlinear $\sigma$ model for the unitary
ensemble in which case all states are known to be
localized\cite{AMW}.
The conclusions of papers \cite{PZ,-,LXSZ}
concerning this problem are still controversial.
It is then probably fair to say that
the question of whether the
Anderson transition is realized
in two-dimensional systems with random magnetic
fluxes remains unsettled and it is a challenging problem to
study the transport properties of such systems.

A two-dimensional system with inhomogeneous
magnetic fields has been
experimentally realized using the type II
superconductors\cite{GBGB}.
Theoretically, a system in random magnetic fields with
zero mean arises in the theory of
composite fermions that describes
the fractional quantum Hall effect
with half-filled Landau levels
\cite{HLR}.
It is therefore a very interesting
problem to clarify not only the
existence of the Anderson transition,
but also the structures of
the wave functions
in such a system.

In this paper we have studied dynamical aspects of the
electron system in static random
magnetic fields by solving the
time-dependent Schr\"{o}dinger
equation numerically.
This type of analysis,
which is usually called the equation-of-motion
method, has been used in the
research of
disordered systems \cite{WW,-,DeRaedt2}.
In particular, the conventional
Anderson model in one- and two-dimensions,
having only a diagonal disorder,
has been studied by De Raedt \cite{DeRaedt2}
based on the exponential product formulae.
However, the forth-order formula used in ref.
\cite{DeRaedt2}
involves
comutators between operators and
becuase of this  it is
difficult to apply it directly to the system
with off-diagonal disorder
such as a system in random magnetic fields.
We therefore adopt here
the new formula for solving the time-dependent
Schr\"{o}dinger equation
proposed by Suzuki \cite{Suzuki}.
This formula has an advantage over
the previous one\cite{DeRaedt2},
because
commutators between operators are not involved.
It is important to note that
owing to this fact
the new formula is easily applicable to
a much wider class of systems
including not only the systems with diagonal disorder
but also those
with off-diagonal disorder \cite{DeRaedt1}.

The quantities we have observed
are the second moment of the wave packet $\langle
r^2 (t) \rangle_c $ defined by
\beq
 \langle r^2 (t) \rangle_c \equiv \langle r^2 (t)\rangle -
 \langle r(t)
 \rangle^2
\eeq
with
\beq
 \langle r^n (t) \rangle \equiv \int
 {\rm d}\Omega r^{d-1} {\rm d}r
 r^n |\psi({\mbox{\boldmath $r$}},t)|^2,
\eeq
where $\psi({\mbox{\boldmath $r$}},t)$ denotes the wave function at time $t$,
and the temporal auto-correlation function $C(t)$ defined by
\beqa
 C(t) & = & \frac{1}{t} \int_0^t P(t') {\rm d}t' \\
 P(t) & \equiv & | \langle \psi(t) | \psi(0)
 \rangle |^2 . \nonumber
\eeqa
Here $P(t)$ denotes the overlap function
between the initial state
$| \psi (0) \rangle $ and
the state $|\psi (t) \rangle$ at
time $t$ and $C(t)$ is its  time average.
These quantities have been calculated
using exact eigenvalues and eigenfunctions of the system
in the cases of the Harper
model\cite{KPG} and the critical state of
the quantum Hall system\cite{HS}.
It has been shown that from the asymptotic
behavior of the auto-correlation
function $C(t)$ we can obtain
information about the fractal
character of the energy levels and
wave functions.
We thus discuss in the present work the fractal character of
the wave function, which is usually observed just at the
critical point or at length-scales much smaller than
the characteristic length
of the system.
Analyses of the fractal character of the wave functions
at the critical energy of the Anderson
transition have been made for
two-dimensional systems in a strong magnetic
field \cite{Aoki1,-,HKS} and
also for three-dimensional systems
\cite{Evangelou,-,GS}. In particular,
the relationship between transport properties
and the fractal structures of the system
is a very interesting problem \cite{BSK}.

The organization of the present paper is as follows.
In the following section,
we explain our method which is
based on the higher-order decomposition
of exponential operators.
The expected asymptotic behaviors of the second
moment
of the wave packet and the
auto-correlation function are also discussed
in this section. In section 3, the above
method is applied to
the
two-dimensional systems in static random magnetic fields
and the numerical results for the
growth of the second moment and the decay
of the auto-correlation are presented.
The final section
is devoted to discussions of the
physical implication of our numerical
results.

\section{Numerical Method}

In order to solve numerically the
time-dependent Schr\"{o}dinger equations,
we have adopted the method based on
the higher-order decomposition
of exponential operators\cite{Suzuki}.
As is seen below, the method is
quite general and easily applicable to
complicated systems described
by the tight-binding Hamiltonian
with inhomogeneous hopping elements.
The basic formula we have used is
the fourth-order decomposition of
exponential operators:
\beq
 \exp [x(A_1 + A_2 + \cdots + A_q)] =
 S(xp) S(x(1-2p)) S(xp) + O(x^5),
 \label{4dec}
\eeq
where
\beq
 S(x) \equiv {\rm e}^{xA_1/2}{\rm e}^{xA_2/2}
 \cdots {\rm e}^{xA_{q-1}/2}
 {\rm e}^{xA_q}{\rm e}^{xA_{q-1}/2} \cdots {\rm e}^{xA_1/2}
\eeq
and the parameter $p$ is given by $p=(2-\sqrt[3]{2})^{-1}$.
Here $A_1, \ldots , A_q$ are arbitrary
operators.

Let us consider, in general, the tight-binding Hamiltonian
on the two-dimensional square lattice
that has nearest neighbor hopping amplitudes
$\{ t_{i,j} ; t_{i,j}^* = t_{j,i} \}$ and the diagonal elements
$\{ \varepsilon_i \}$:
\beq
 H = -\sum_{<i,j>} t_{i,j}
 C_i^{\dagger} C_j + \sum_i \varepsilon_i
 C_i^{\dagger} C_i,
\eeq
where $C_i^{\dagger}(C_i)$ denotes
a creation (annihilation) operator
of an electron at the site $i$.
We then decompose this Hamiltonian
into five parts (see figure 1),
namely,
\beqa
 H & = & \sum_{n=1}^5 H_n , \\
 H_1 & \equiv & \sum_{r_x \in {\rm odd}} \sum_{r_y}
 t_{\vec{r}+\hat{x},\vec{r}} C_{\vec{r}+\hat{x}}^{\dagger}
 C_{\vec{r}} + {\rm h. c.} \nonumber , \\
 H_2 & \equiv & \sum_{r_x \in {\rm even}} \sum_{r_y}
 t_{\vec{r}+\hat{x},\vec{r}} C_{\vec{r}+\hat{x}}^{\dagger}
 C_{\vec{r}} + {\rm h. c.} \nonumber , \\
 H_3 & \equiv & \sum_{r_x} \sum_{r_y \in {\rm odd}}
 t_{\vec{r}+\hat{y},\vec{r}} C_{\vec{r}+\hat{y}}^{\dagger}
 C_{\vec{r}} + {\rm h. c.} \nonumber \\
 H_4 & \equiv & \sum_{r_x} \sum_{r_y \in {\rm even}}
 t_{\vec{r}+\hat{y},\vec{r}} C_{\vec{r}+\hat{y}}^{\dagger}
 C_{\vec{r}} + {\rm h. c.} \nonumber , \\
 H_5 & \equiv & \sum_{\vec{r}} \varepsilon_{\vec{r}}
 C_{\vec{r}}^{\dagger}
 C_{\vec{r}} .\label{decom}
\eeqa
In equations (\ref{decom}), $\hat{x}(\hat{y})$
denotes the unit vector
in the $x(y)-$direction and $r_x(r_y) \in {\mbox{\boldmath $Z$}}$ the $x(y)$
component
of the position vector $\vec{r}$ of the sites.
All the length-scales are measured
in units of the lattice constant $a_0$.
The Hamiltonian $H_1$, for instance,
describes the hoppings between
the sites whose positions are
$\vec{r}$ and $\vec{r}+ \hat{x}$ for
$r_x=1,3,5,\ldots $ and $r_y = 1,2,3,
\ldots $.
Note that each Hamiltonian $H_n$ consists of operators
which commute with each other, and
hence we can obtain the analytical
expressions for $\exp(-{\rm i}H_n t/\hbar)$
by diagonalizing two by two matrices.
According to the formula (\ref{4dec}),
it is easy to see that
the state vector $|\psi(t + \delta t) \rangle$
at time $t+\delta t$  can be
obtained as
\beqa
 | \psi(t+ \delta t) \rangle & = &
 \exp(-{\rm i}\delta t H /\hbar) |\psi(t)\rangle \nonumber \\
 & = &
 S_2(-{\rm i}\delta t p /\hbar)
 S_2(-{\rm i}\delta t(1-2p) /\hbar)
 S_2(-{\rm i}\delta t
 p /\hbar) | \psi (t) \rangle \nonumber \\
 & & + O(\delta t^5)
\eeqa
with
\beq
 S_2(x) \equiv {\rm e}^{xH_1/2}{\rm e}^{xH_2/2}\cdots
 {\rm e}^{xH_4/2}
        {\rm e}^{xH_5}{\rm e}^{xH_4/2} \cdots
        {\rm e}^{xH_2/2}{\rm e}^{xH_1/2} .
\eeq
It should be noted that in the present approach the unitarity
of the process is obviously
preserved and that we have high enough accuracy to perform
large time-scale simulations.
It is important to note that in this method
we do not need to diagonalize the whole
Hamiltonian. We can thus simulate larger systems,
for example a 499 by 499
square lattice, with high precision, which
is crucial if we are to extract meaningful properties
from the numerical data
for random systems.

In order to consider the time evolution of
the wave packet with  fixed
energy $E$, we have carried out a numerical diagonalization  of
a subsystem
$H_{N_0}$
whose size is $N_0$ by $N_0$, located
at the center of the whole system and
have chosen an eigenstate of $H_{N_0}$ with
eigenvalue $E_{N_0} \approx E$ as
the initial wave function.

Next, let us consider the asymptotic
behaviors of the auto-correlation
function $C(t)$ and the second moment
of the wave packet $\langle r^2 \rangle_c$
in several regions, namely, the metallic,
the localized (insulating) and
the critical regions. First, in the localized region
it is clear that
the second moment remains finite
$\langle r^2(t) \rangle_c \rightarrow
d(d+1)\xi^2/4$ in the limit as $ t \rightarrow \infty $,
where $\xi$ denotes
the localization length
\cite{DeRaedt2} that describes the exponential-decay
of the wave function as $|\psi(r)| \propto \exp[-r/\xi ]$.
Obviously, the quantity $C(t)$ also goes to
a nonzero value in the same limit.
In contrast, in the metallic
region it is expected that the second moment grows in proportion to
time $t$:
\beq
 \langle r^2 \rangle_c = 2dDt \label{diffc}
\eeq
and that the auto-correlation $C(t)$ decreases as
\beq
 C(t) \propto t^{-d/2}.
\eeq
Here, the coefficient $D$ denotes
the diffusion coefficient and $d$
the dimensionality of the system.

In the critical region where
the wave function has a fractal structure,
a
non-conventional behavior of
the auto-correlation function $C(t)$ has been
observed\cite{KPG,HS}. In such a case, the
auto-correlation function decays
in proportion to $t^{-\alpha}$ for
$t \gg 1$ with $\alpha \neq d/2$
reflecting the fractal character
of the energy spectrum \cite{KPG}
and of the wave function \cite{HS}.
It has been shown that
the exponent $\alpha$ is related
to the generalized dimension
$\tilde{D}_2$ of
the energy spectrum
by $\alpha = \tilde{D}_2 $ \cite{KPG}, and, further, that
in the case of the quantum Hall system it is also related to
the generalized fractal dimension $D_2$ \cite{PJ} of the
wave function
by $\alpha = D_2/2$ \cite{HS}.
The generalized dimension
$\tilde{D}_2$ of
the energy spectrum
is defined as the scaling exponent for the correlation of
energy levels:
\beq
 \gamma (l) \equiv \int {\rm d}\mu_{\psi (0)}( \omega )
 \int_{\omega - l}^{\omega + l} {\rm d}\mu_{\psi (0)}(\omega ')
 \propto l^{\tilde{D}_2} \quad{\rm as}\quad l \rightarrow 0 ,
\eeq
where $\mu_{\psi(0)}(\omega)$ denotes the spectral measure with
respect to an initial state $\psi(0)$ \cite{KPG}.
The generalized fractal dimension $D_2$ of the
wave function is, on the other hand, defined by:
\beq
 P_2 (\lambda , E)
 \equiv \sum_i \bigg( \sum_{r \in \Omega_i ( \lambda )}
 | \psi (r,E) |^2 \bigg)^2 \propto
 \lambda^{D_2} \quad {\rm as} \quad
 \lambda \rightarrow 0 ,
\eeq
where $\{ \Omega_i(\lambda ) \}$ denote $ \lambda N
\times \lambda N $ boxes
that cover the whole system ($N \times N$)
\cite{HS}.
On the basis of these
relations,
the non-conventional behavior of the auto-correlation
function $C(t)$ can be considered as  evidence for
fractal structure in the system.
Since the multifractal behavior has been demonstrated not
only at the critical point, but also for length-scales
smaller than the characteristic
length, such as the localization length
in the localized region \cite{SG},
a similar type of
non-conventional behavior of $C(t)$ is
expected if we consider
length-scales much smaller than the characteristic length
of the system.

\section{Model and Numerical Results}

The two-dimensional system
with random magnetic fluxes is described
by the tight-binding Hamiltonian:
\beq
 H = - \sum_{<i,j>} V_{i,j} C_i^{\dagger} C_j ,
\eeq
where
\beq
 V_{i,j} = V \exp [ {\rm i} \theta_{i,j}].
\eeq
The phases $\{ \theta_{i,j} \}$ are related
to the magnetic fluxes
$\{ \phi_i \}$
through the relation:
\beq
 \theta_{i+\hat{x},i} +
 \theta_{i+\hat{x}+\hat{y},i+\hat{x}}
 +\theta_{i  +\hat{y}, i+ \hat{x}+\hat{y}} +
 \theta_{i,i+\hat{y}}
 = -2\pi \phi_i/\phi_0,
\eeq
where $\phi_i$ and $\phi_0 \equiv hc/|e|$ denote
the magnetic flux through the plaquette
$(i,i+\hat{x},i+\hat{x}+\hat{y},i+\hat{y})$ (figure 2) and
the unit flux, respectively.
It should be noted here that
we have the gauge degrees of freedom
in determining the phases
$\{ \theta_{i,j} \}$ for given fluxes
$\{ \phi_i \}$.
The magnetic fluxes through each
plaquettes of the square lattice
are distributed
uniformly in the interval $[ -\phi_0 /2, \phi_0/2 ]$, and the
fluxes are assumed to have no spatial correlation, i.e.
\beq
 \langle \phi_i \phi_j \rangle_{\rm av}
 = \frac{1}{12}\phi_0^2 \delta_{ij}.
\eeq
Here the angular bracket $\langle \quad \rangle_{\rm av}$
denotes the average over the distribution
of fluxes.

We have calculated the auto-correlation
function $C(t)$ and the
second moment of the wave packet
$\langle r^2 \rangle_c$ at various
energies. The sizes of the systems
are 499 by 499 for the energies
near the band center and
199 by 199 for the energy near the band edge.
We have carried out an exact diagonalization for
the 21 by 21 subsystem at the center of the system and
taken the eigenfunction
of the subsystem whose eigenvalue is closest to
the given energy $E$ as the initial wave packet.
By this procedure we can
simulate the diffusion of the wave
packet whose energy is approximately
equal to $E$.
The single time step $\delta t$ is
taken to be $0.2 (\hbar/V)$
in the simulation. With this choice of $\delta t$,
fluctuations of the
expectation value of the Hamiltonian
\beq
 \langle H \rangle \equiv \langle
 \psi (t) | H | \psi (t) \rangle
\eeq
can be safely neglected \cite{DeRaedt2} throughout
our simulations
($ t \leq 8000 (\hbar/V)$).
In the actual simulation, the quantities
$\langle r^2 \rangle_c$ and $\log C(t)$ are
averaged over five realizations
of random magnetic flux distribution.

Let us first discuss the diffusion near the band edge.
The states near the band edge
are expected to be localized \cite{PZ,SN}.
Our results for the energy
$\langle H \rangle \sim -3.35 V $ are shown
in figures 3(a) and 3(b),
which is based on a calculation of five
configurations of random magnetic fields.
The second moment of the wave packet
$\langle r^2 \rangle_c$ seems to have
a finite value in the limit $t \rightarrow \infty$.
The localization length can be
roughly estimated as $\xi \sim 14$,
which is consistent with
the value obtained from the finite-size scaling
analysis of ref. \cite{SN}.
It may not be clear from the behavior of the
second moment of the wave packet
alone whether the states are indeed
localized or not.
The behavior of the auto-correlation $C(t)$, however,
clearly suggests that the state at
time $t$ has a nonzero overlap integral
with the initial state
in the limit $t \rightarrow \infty$ and
thus the states are localized in this energy region.

Next, we discuss the diffusion near the band center. The
results for the energies $\langle H \rangle /V$ $\sim$ 0, $-1$,
$-2$, $-2.6$ are presented in
figures 4, 5, 6 and 7, respectively. The growth of the
second moment of the wave
packets turns out to be  diffusive; the numerical
value of the exponent $a$ for
the power-law growth defined in
\beq
 \langle r^2 (t) \rangle_c \propto t^a
\eeq
is very close to 1. It should
be noted here that, in the critical case of the Harper model,
the exponent $a$ has been estimated to be
about 0.97 \cite{HA,HK}, which is also close to $1$ \cite{WA}.
The exponent $a$ has been also estimated to be $1$ for the
critical states of the quantum Hall system \cite{HS}.
By assuming a linear dependence of
$\langle r^2 (t) \rangle$ with
respect to $t$, we can
evaluate the diffusion coefficient $D$ defined in
eq.(\ref{diffc}), which is related
to the conductance of the system.
The obtained values of the
diffusion coefficients for energies $-2.6$,
$-2$, $-1$ and $0$ are consistent
with those obtained using the Landauer
formula \cite{OSO}.

On the other hand, the exponents
$\alpha$ for the power-law decay of the
auto-correlation function
show deviations from the value $d/2=1$
in the same
energy region. For the energy region $-2.6 \leq E \leq 0 $,
the exponent $\alpha$
is estimated to vary from
$0.9$ to $ 0.88$, which is smaller than $1$.
These deviations suggest the presence of fractal properties
of the energy spectrum and the wave function, just as observed
in the cases of the
Harper model and the quantum Hall system \cite{KPG,HS}.
The deviations themselves are smaller than that
obtained for the critical
case of the quantum Hall system \cite{HS},
suggesting a larger fractal
dimensionality of the present system.

\section{Discussions}

We have performed numerical simulations for the
diffusion of electrons in static random magnetic fields
using a new formula
for the decomposition of exponential operators
\cite{Suzuki}.
The systems we have
treated are much larger than those in previous
studies of this kind \cite{DeRaedt2,HS}.
The present method
reproduces the results previously found from the
finite-size scaling analyses, namely
that the states near the band edge are localized.
On the other hand,
away from the band edge, the second moment of the
wave packets increases linearly in time, which suggests
that the conductance is non-zero in this region.
If the system is described
by the conventional diffusion equation,
the exponent $\alpha$ should be
$d/2=1$ and in that case no fractal structure is present.
However,
the obtained values of
the exponents $\alpha$ of
the power-law decay of the auto-correlation
function are smaller than $d/2=1$.
It is expected from these deviations of the values of $\alpha$
from $1$ that
some kind of fractal structure  exists
in the energy
spectrum as well as in the eigenfunctions of the present system.
This type of behavior,
namely the linear increase of $\langle r^2 \rangle_c$ in time
and the non-conventional
exponents for the power-law decay
of the auto-correlation function,
has been observed  in the critical case of the
Harper model and the critical states
in the quantum Hall system \cite{KPG,HS}.
It may therefore be natural to
suppose that the states away from the band
edge
in the present system with random magnetic fluxes
are also critical.

It has been shown that, by assuming  conformal invariance,
the exponent $\eta$ for the
power-law decay of the critical
wave function in two-dimensional systems
can be related to the exponential decay-length
of the Green function
in quasi-one dimensional strip systems
\cite{Cardy,PS}. The exponent $\eta$,
which describes the power-law
decay of the density correlation by
$\langle |\psi(r)|^2 |\psi(0)|^2 \rangle_{\rm av}
\sim |r|^{-2\eta}$ in
two dimensions,
is expressed as  \cite{PS}
\beq
 \eta = \frac{1}{\pi \Lambda_c}, \label{form1}
\eeq
where $\Lambda_c$ denotes
the renormalized localization length, defined by
\beq
 \Lambda_c \equiv \lim_{M \rightarrow \infty}
 \xi_M (E) /M. \label{form2}
\eeq
Here $\xi_M (E)$ denotes the
localization length along a strip of width $M$
with energy $E$ corresponding
to the critical states in infinite
two-dimensional systems
\cite{MK}.
The fractal dimensionality $d^*$ of the
wave function is then obtained
from the exponent $\eta$ as \cite{PS}
\beq
 d^* = d - 2\eta . \label{form3}
\eeq
In this way, the fractal dimensionality $d^*$ of the wave
function in a two-dimensional system
at its critical energy,
or at the length-scales much smaller than the
characteristic length,
can thus be evaluated from the
scaling behavior of the
localization length along the quasi-one-dimensional
systems.
Making use of this theory \cite{PS},
we have performed independently
the calculation of $\xi_M$ for a few random
field configurations in strips whose sizes are
$M \times L$ with $L =10^5$ and $M=8$, $12$,
$16$, $20$, $34$, $40$ and $50$ in units of
the lattice spacing $a_0$
(figure 8) and
have estimated the exponent
$\eta$ through the formulae (\ref{form1}) and
(\ref{form2}).
The values of $d^*$ obtained by the
formula (\ref{form3}) are then
compared with the generalized
fractal dimension $D_2$ obtained from the
exponent $\alpha$ by assuming
the relation $D_2 = 2\alpha$ \cite{HS}
(Table 1).
It is  rather remarkable that the values
of $d^*$ and $D_2$ coincide fairly well with each other.

We have thus argued,
based on our numerical results, that there is
evidence for
fractal structure in the energy spectrum and the eigenfunction
in the region away from the band edge. However,
we admit that the fractal character of the
present system turns out to be not
so strong compared with the case
of the critical states in
the quantum Hall system \cite{HS}. At the
band center, in particular,
we cannot exclude the possibility that
there is no fractal structure,
because the value of the exponent
$\alpha$ turns out to be rather close to $1$.
Since the fractal structure is observed when
the characteristic length is much larger than
the size of the system, then precisely speaking, our results
do not distinguish the critical states from the
localized states with
extremely large localization length.
Simulations in much larger systems and
the investigation of the probability
distributions of $\langle r^2 \rangle_c$ and $C(t)$
will be needed in order to obtain
more detailed information about the
thermodynamic properties of this system.

Finally, we emphasize that
the formula we have used in this work is quite general.
It would be very interesting to study
many other dynamical problems,
such as the anomalous diffusion \cite{SSV} and
the mesoscopic dynamical echo \cite{PAEI} by making use of the
present method.

\section*{Acknowledgments}

The authors would like to thank Profs. Y.~Ono, J.~Kondo,
B.~Kramer, M.~Kohmoto and M.~Takahashi
for fruitful discussions.
They also thank Dr. S.M.~Manning for reading the manuscript.
Numerical calculations were performed on HITAC
S-3800 of the Computer Center, University of Tokyo. This work
was partly
supported by the Grant-in-Aid No.~06740316, No.~04231105
and
No.~05854024,
from the Ministry of Education,
Science and Culture, Japan. One of the authors (T.~K.) thanks
the F\={u}jukai Foundation for a scholarship.

\clearpage

\begin{table}

\begin{tabular*}{15cm}
 {c@{\extracolsep{\fill}}c
 @{\extracolsep{\fill}}c@{\extracolsep{\fill}}c}
 \hline
 $E$ & $\Lambda_c$
 & $d^* = d -2\eta $ & $D_2 = 2 \alpha$ \\ \hline \hline
 $0$ & $3.04 \pm 0.04$ & $1.791\pm 0.002$ &
 $1.8 \pm 0.06$ \\ \hline
 $-1$ & $3.00 \pm 0.08$ & $1.788\pm 0.006$ &
 $1.8 \pm 0.06$ \\ \hline
 $-2$ & $2.73 \pm 0.07$ & $1.766\pm 0.006$ &
 $1.8 \pm 0.03$ \\ \hline
 $-2.6$ & $2.34 \pm 0.09$ & $1.73\pm 0.01$ &
 $1.76 \pm 0.02$ \\
 \hline
\end{tabular*}
\caption{Fractal dimensions obtained from
the exponents $\eta$ and
those from the exponents $\alpha$ for
$E$ = $0$, $-1$, $-2$, $-2.6$.
The values $\Lambda_c$ are estimated using
$M \times L$ strips with $8 \leq M \leq 50$
and $L=10^5$.}

\end{table}

\clearpage

\begin{figcaption}
\item The decomposition of the Hamiltonian.
      The thin solid, thick solid,
      thin dashed and thick dashed lines
      correspond to the Hamiltonian
      $H_1$, $H_2$, $H_3$ and $H_4$ in
      equation (\ref{decom}), respectively.

\item The plaquette $(i, i + \hat{x},
      i+\hat{x}+\hat{y}, i+\hat{y})$ and the
      flux $\phi_i$ through it.

\item (a) The growth of the second
       moment $ \langle r^2 (t) \rangle_c$
       of the wave packet
       and (b) the time-dependence of
       the auto-correlation function $C(t)$
       for the energy $\langle H \rangle$
       $\simeq$ $-3.35 (-3.35 \pm 0.01)V$.
       The bars around the data points in
       figures 3, 4, 5, 6 and 7 indicate the
       fluctuations of the
       mean values with respect to five
       realizations of random magnetic field
       configurations.

\item (a) The growth of the second moment
       $ \langle r^2 (t) \rangle_c$
       of the wave packet
       and (b) the time-dependence of the
       auto-correlation function $C(t)$
       for the energy $\langle H \rangle$ $\simeq$
       $0(\pm 1.0 \times 10^{-4})V$.
       The solid line in (a) corresponds to
       $D = 1.59 a_0^2 V/\hbar$.
       The exponent $\alpha$
       is estimated to be $0.90 \pm 0.03$. For comparison,
       the conventional decay of
       $C(t) \propto t^{-1}$ is also shown by
       the dashed lines in figures
       4(b), 5(b), 6(b) and 7(b).

\item (a) The growth of the second
       moment $ \langle r^2 (t) \rangle_c$
       of the wave packet
       and (b) the time-dependence of the
       auto-correlation function $C(t)$
       for the energy $\langle H \rangle$ $\simeq$
       $-1 (-1.01 \pm 0.02) V$.
       The solid line in (a) corresponds to
       $D = 1.54 a_0^2 V/\hbar$.
       The exponent $\alpha$
       is estimated to be  $0.90 \pm 0.03$.

\item (a) The growth of the second moment
       $ \langle r^2 (t) \rangle_c$
       of the wave packet
       and (b) the time-dependence of the
       auto-correlation function $C(t)$
       for the energy $\langle H \rangle$ $\simeq$
       $-2 (-1.99 \pm 0.01)V$. The solid line in (a)
       corresponds to
       $D = 1.33 a_0^2 V/\hbar$.
       The exponent $\alpha$
       is estimated to be $0.90 \pm 0.014$.

\item (a) The growth of the second moment
       $ \langle r^2 (t) \rangle_c$
       of the wave packet
       and (b) the time-dependence of
       the auto-correlation function $C(t)$
       for the energy $\langle H \rangle$
       $\simeq$ $-2.6 (-2.61 \pm 0.01)V$.
       The solid line in (a) corresponds to
       $D = 0.98 a_0^2 V/\hbar$.
       The exponent $\alpha$
       is estimated to be $0.88 \pm 0.01$.

\item  Energy dependence of the exponential
       decay length $\xi_M$ divided by
       the width of the strip $M$ \cite{SN,MK}.
       The circle, square, diamond,
       triangle, bullet, painted square and
       painted diamond correspond to
       $M=8.12,16,20,30,40$ and 50
       respectively.
       The length of the strip $L$ is $10^5$ and
       the average is performed
       over more than two random flux configurations.
       Due to the exact particle-hole symmetry,
       only the negative energy
       region is plotted.
\end{figcaption}

\end{document}